\documentclass[12pt]{article}
\usepackage{epsf,amsfonts}

\epsfverbosetrue
\def\goth{\mathfrak}          
\def\double{\mathbb}         
\def\ccal{\cal}           

\def\cc{{\double C}}     
       
\def\rr{{\double R}}

\def\aa{{\cal A}}

\def\dd{{\cal D}} 
\def\gg{{\goth g}}        
\def\hh{{\cal H}}
\def\hhh{{{\double H}}}   

\def\mm{{{\ccal M}}}

\def\aa{{\cal A}}
\def\dd{{\cal D}} 
\def\hh{{\cal H}}

\def\t{{\rm tr}}

\def\dee{\,\hbox{\rm D}}
\def\de{\,\hbox{\rm d}}

\def\lb{\left[} 
\def\rb{\right]}

\def\ot{\otimes}
\def\op{\oplus}

\def\bb{\begin{eqnarray}}
\def\ee{\end{eqnarray}}
\def\eee{\nonumber\end{eqnarray}}
\def\pp{\pmatrix}
\def\qq{\quad}

\begin{document}

\hsize 17truecm
\vsize 24truecm
\font\twelve=cmbx10 at 13pt
\font\eightrm=cmr8
\baselineskip 18pt

\begin{titlepage}

\centerline{\twelve CENTRE DE PHYSIQUE THEORIQUE}
\centerline{\twelve CNRS - Luminy, Case 907}
\centerline{\twelve 13288 Marseille Cedex}
\vskip 4truecm

\centerline{\twelve Non-commutative Geometry
Beyond the Standard Model }

\bigskip

\begin{center}
{\bf Igor PRIS}
\footnote{ and Universit\'e de Provence,\qq
pris@cpt.univ-mrs.fr} \\
\bf Thomas SCH\"UCKER 
\footnote{ and Universit\'e de Provence,
\qq schucker@cpt.univ-mrs.fr} \\

\end{center}

\vskip 2truecm
\leftskip=1cm
\rightskip=1cm
\centerline{\bf Abstract} 

\medskip

A natural extension of the standard model within
non-commutative geometry is presented. The
geometry determines its Higgs sector. This
determination is fuzzy, but precise enough to be
incompatible with experiment.

\vskip 1truecm
PACS-92: 11.15 Gauge field theories\\ 
\indent
MSC-91: 81E13 Yang-Mills and other gauge theories 
 
\vskip 2truecm

\noindent march 1996
\vskip 1truecm
\noindent CPT-96/P.3335\\
\noindent hep-th/9604115
 \vskip1truecm
 \end{titlepage}

The standard model of electro-weak and strong
interactions is a mediocre element of the huge set of
Yang-Mills-Higgs theories. Analyzing neighboring
theories in this set has essentially two motivations,
Lagrange's principle of variation to see whether
there is a better theory in the vicinity and an
assessment of experimental deviations from the
standard model. 

Non-commutative
geometry allows us --- in some cases --- to understand
the Higgs field as a magnetic field of a Yang-Mills
gauge field. There are essentially three approaches to
this idea. The approach due to Dubois-Violette, Kerner
and Madore \cite{DKM} is the most
restrictive one. It applies to Yang-Mills theories with
unbroken parity. The approach due to  Coquereaux,
Esposito-Far\`ese and Vaillant \cite{CEV},
on the other hand, is so general that we do not have a
model building kit. In the following, we shall stick to
Connes' approach \cite{C}, that to our taste \cite{RV}
has the most appealing geometrical motivation. This
approach is restricted to a tiny set of Yang-Mills
theories with Dirac fermions. That the standard
model belongs to this set, is a miracle to us. In order to
appreciate it, we look for extensions of the standard
model within Connes' frame. These are not easy to
find. Left-right symmetric models and grand unified
theories do not fit into Connes' frame \cite{IS}. No
realistic supersymmetric model has been found so far
\cite{KW}. 

The mild extension of the standard model,
that we would like to discuss here is motivated
\cite{TR} by the quantum group $SU(2)_j$ with $j$ a
cubic root of 1. Non-commutative geometry is the
geometry of spaces where points are excluded by an
uncertainty relation. The phase space in
quantum mechanics is the first example of a
non-commutative geometry. Today, the word
`quantum' is so overused that we prefer Madore's
terminology. He calls these spaces {\it fuzzy} and his
fuzzy sphere is a most instructive example \cite{FU}.
According to Connes, the quantum group is to a fuzzy
space what the Lie group is to a manifold. So far the
quantum group of the standard model is unknown, but
the hope is that this quantum group will explain the
{\it fuzzy} mass relation for the Higgs mass \cite{IKS}, 
\bb  m_H^2\ =\ 3\, {m_t}^2-
m_W^2\,\left(1+\frac{{g_2}^{-2}}
{{g_1}^{-2}-\frac{1}{6}{g_3}^{-2}}\right)
\ +\
O\left(\frac{m^4_\tau}{m^2_t}\right)\ee
which appears if we want to fit the standard model
into Connes' frame. $SU(2)_j$ co-acts on the
associative algebra $M_2(\cc)\op M_1(\cc)\op
M_3(\cc)$ which extends mildly the algebra of the
standard model, $\hhh\op \cc\op M_3(\cc)$.

\section{Input}

The input of Connes' model building kit is a spectral
triple $(\aa,\hh,\dd)$ and a non-commutative
coupling. $\aa$ is an associative involution
algebra with unit. Its group of unitaries,
\bb G:=\left\{g\in \aa\ |\ g^*g=gg^*=1\right\} ,\ee
or a subgroup thereof will be the group of gauge
transformations. $\hh$ is a Hilbert space that
carries a faithful representation $\rho$ of $\aa$. The
Hilbert space is supposed to decompose into four pieces,
\bb \hh=\hh_L\op\hh_R\op\hh_L^c\op\hh_R^c,\ee
containing the left- and
right-handed fermions and anti-fermions,
\bb \rho=\pp{\rho_L&0&0&0\cr 0&\rho_R&0&0\cr
0&0&\bar\rho_L^c&0\cr 0&0&0&\bar\rho_R^c}.\ee
 $\dd$ is the Dirac operator,
an odd, selfadjoint operator on $\hh$. $\dd$ contains
the fermionic mass matrix $\mm$,
\bb\dd=\pp{
0&\mm&0&0\cr 
\mm^*&0&0&0\cr 
0&0&0&\overline{\mm}\cr 
0&0&\overline{\mm^*}&0},\ee
with respect to the above decomposition of $\hh$.  The
non-commutative coupling
$z$ parameterizes invariant scalar products and
therefore generalizes the Yang-Mills gauge couplings.
$z$ is an even, positive operator on
$\hh$, 
\bb z=\pp{z_L&0&0&0\cr 0& z_R&0&0\cr 
0&0&z_L^c&0\cr 0&0&0&z_R^c},\ee
that commutes with $\rho$ and $\dd$.

For the standard model, the spectral triple and the
coupling are:
\bb \aa=\hhh\op \cc\op M_3(\cc)\ni (a,b,c),\ee
$\hhh$ denoting the quaternions,
\bb\hh_L&=&
\left(\cc^2\ot\cc^N\ot\cc^3\right)\ \op\ 
\left(\cc^2\ot\cc^N\ot\cc\right),\\
\hh_R&=&\left((\cc\op\cc)\ot\cc^N\ot\cc^3\right)\ 
\op\ \left(\cc\ot\cc^N\ot\cc\right).\ee
 In each summand, the first factor
denotes weak isospin doublets or singlets, the second -
$N$ generations, $N=3$, and the third denotes colour
triplets or singlets. 
\hfil\break\noindent
Let us choose the following basis
of 
$\hh=\cc^{90}$: 
\bb
& \pp{u\cr d}_L,\ \pp{c\cr s}_L,\ \pp{t\cr b}_L,\ 
\pp{\nu_e\cr e}_L,\ \pp{\nu_\mu\cr\mu}_L,\ 
\pp{\nu_\tau\cr\tau}_L;&\cr \cr 
&\matrix{u_R,\cr d_R,}\qq \matrix{c_R,\cr s_R,}\qq
\matrix{t_R,\cr b_R,}\qq  e_R,\qq \mu_R,\qq 
\tau_R;&\cr  \cr 
& \pp{u\cr d}^c_L,\ \pp{c\cr s}_L^c,\ 
\pp{t\cr b}_L^c,\ 
\pp{\nu_e\cr e}_L^c,\ \pp{\nu_\mu\cr\mu}_L^c,\ 
\pp{\nu_\tau\cr\tau}_L^c;&\cr\cr  
&\matrix{u_R^c,\cr d_R^c,}\qq 
\matrix{c_R^c,\cr s_R^c,}\qq
\matrix{t_R^c,\cr b_R^c,}\qq  e_R^c,\qq \mu_R^c,\qq 
\tau_R^c,&\ee
\bb \rho_L(a)=\pp{
a\ot 1_N\ot 1_3&0\cr
0&a\ot 1_N},&&
\rho_R(b)=\pp{
B\ot 1_N\ot 1_3&0\cr
0&\bar
b1_N},\qq
B:=\pp{b&0\cr 0&\bar b},
\cr  \cr \cr 
  \rho_L^c(b,c)=\pp{
1_2\ot 1_N\ot c&0\cr
0&\bar b1_2\ot 1_N},&&\rho_R^c(b,c)=\pp{
1_2\ot 1_N\ot c&0\cr
0&\bar b1_N},   \ee
\bb\mm=\pp{
\pp{M_u\ot1_3&0\cr 0&M_d\ot 1_3}&0\cr
0&\pp{0\cr M_e}},\ee
with
\bb M_u:=\pp{
m_u&0&0\cr
0&m_c&0\cr
0&0&m_t},\qq M_d:= C_{KM}\pp{
m_d&0&0\cr
0&m_s&0\cr
0&0&m_b},\qq M_e:=\pp{
m_e&0&0\cr
0&m_\mu&0\cr
0&0&m_\tau}.\ee
 All indicated fermion masses are supposed positive and
different. The
Cabibbo-Kobayashi-Maskawa matrix  
\bb C_{KM}:=\pp{V_{ud}&V_{us}&V_{ub}\cr 
V_{cd}&V_{cs}&V_{cb}\cr 
V_{td}&V_{ts}&V_{tb}}\ee
is supposed non-degenerate in the sense that there is
no simultaneous mass and weak interaction
eigenstate. The coupling $z$ involves six positive
numbers $x,\ y_1,\ y_2,\ y_3,$ $ \tilde x,\ \tilde y$,
\bb z_L=\pp{
x/3\,1_2\ot 1_N\ot 1_3&0\cr  0&1_2\ot y},&&
z_R=\pp{ 
x/3\,1_2\ot 1_N\ot 1_3&0\cr 
0&y},\cr \cr \cr 
z_L^c:=\pp{
\tilde x/3\,1_2\ot 1_N\ot 1_3&0&\cr  0&\tilde
y/3\,1_2\ot  1_3},&&z_R^c=\pp{ 
\tilde x/3\,1_2\ot 1_N\ot 1_3&0\cr 
0&\tilde y/3\,1_3},\ee with 
\bb y:=\pp{ 
y_1&0&0\cr 
0&y_2&0\cr 
0&0&y_3}.\ee

Note that on the level of algebra representation, we
have an asymmetry between particles and
anti-particles, the former are subject to weak, the
latter to strong interactions. This asymmetry is lifted
later at the level of the Lie algebra representation
$\tilde \rho$.
 
The proposed
extension of the standard model is mild, we extend the
quaternions to arbitrary complex
$2\times 2$ matrices, $a\in M_2(\cc)$,
\bb \aa= M_2(\cc)\op M_1(\cc)\op M_3(\cc).\ee
 All other input
items are unchanged. Nevertheless, compared to the
standard model, the calculations will turn out to be
quite different and longer.

\section{Turning the crank}

We shall organize our calculations according to the
theorem of \cite{SZ} with invariant scalar product
${\rm Re}\ \t [\rho(\cdot)^*\rho(\cdot)\,z]$.
Our first task is to compute the 1-forms. 
$\rho^c$ being vectorlike does not produce 1-forms
and momentarily we may restrict ourselves to
$\hh_L\op\hh_R.$ A general 1-form is a sum of terms 
\bb   
\pi((a_0,b_0)\delta(a_1,b_1)) =-i \pmatrix
{0&\rho_L(a_0)\left(\mm\rho_R(b_1)-\rho_L(a_1)
\mm\right)\cr \rho_R(b_0)\left(\mm^*\rho_L(a_1)-
\rho_R(b_1) \mm^*\right)&0}\ee  
 and the 
vector space of 1-forms is
\bb   \Omega_\dd^1\aa = \left\{i\pmatrix
{0&\rho_L(h)\mm\cr \mm^*\rho_L(\tilde h^*)&0},\
h,\tilde h\in M_2(\cc)\right\}.\ee
Our basic variable, the `Higgs', is an anti-Hermitian
1-form
\bb H=  i\pmatrix
{0&\rho_L(h)\mm\cr \mm^*\rho_L(h^*)&0},\qq
h=\pp{h_{11}& h_{12}\cr  h_{21}&  h_{22}}
\in
M_2(\cc).\ee
It is parameterized by {\it two} isospin doublets
\bb h_1= \pp{h_{11}\cr h_{21}},\qq h_2=\pp{
h_{12}\cr 
 h_{22}}.\ee

Our next task is to compute the 2-forms. The junk in
degree two is:
\bb J^2=\left\{i\pp{j\ot\Delta&0\cr 0&0},\qq
j\in M_2(\cc)
\right\}\ee
with
\bb\Delta:=\frac{1}{2}\pp{\left(M_uM_u^*-M_dM_d^*
\right)\ot 1_3&0\cr 
0&-M_eM_e^*}.\ee
With respect to the scalar product
 the 2-forms   are written as
\bb
\Omega_\dd^2\aa=\pi(\Omega^2\aa)/
J^2=\left\{\pp{
\tilde c\ot \Sigma' &0\cr 
0&\mm^*\rho_L(c)\mm},\qq 
\tilde c,c\in M_2(\cc)\right\}\ee
with
\bb\Sigma'=\Sigma - \eta \frac{\t (\Sigma 
\Delta z_\ell )}{\t (\Delta^2 z_\ell )} \Delta,\ee
\bb z_\ell=\pp{
(x/3)1_N\ot 1_3&0\cr  0& y},&&
 \Sigma:=\frac{1}{2}\pp{\left(M_uM_u^*+M_dM_d^*
\right)\ot 1_3&0\cr 
0&M_eM_e^*};\ee
 \bb \eta=1&{\rm
for\ the \ case} & \tilde{c},c\in
M_2(\cc),   \\
  \eta =0&{\rm
for\ the \ case} &\tilde{c},c\in  \hhh   
\qquad{\rm (the \ standard\ model)}.   \ee

For the  differential $\delta:
\Omega_\dd^1\aa\longrightarrow
\Omega_\dd^2\aa$ we have:
\bb
i\pmatrix
{0&\rho_L(h)\mm\cr \mm^*\rho_L(\tilde h^*)&0}
 &\longmapsto& \pp{
(h+\tilde h^*)\ot \Sigma'&0\cr 
0&\mm^*\rho_L(h+\tilde h^*)\mm}.\ee    

Now we can  compute the curvature 
\bb C:=\delta H+H^2=
\pp{\left(h+h^*-hh^*\right)\ot\Sigma'&0
\cr 
0&M^*\rho_L(h+h^*-h^*h)M},\ee     
where 
\begin{eqnarray}
 \Phi:=H-i\pp{0&\mm\cr \mm^*&0}
=:i\pmatrix
{0&\rho_L(\varphi)\mm\cr
\mm^*\rho_L(\varphi^*)&0},\ \varphi=h-1,  
&&\varphi=(\varphi_1,\varphi_2),  \nonumber \\
    &&
\varphi,  h    \in  M_2(\cc),
\end{eqnarray}
is the homogeneous scalar variable.

In order to compute the Higgs potential \cite{SZ}
we must return to $\cc^{90}$,
\bb V:={\rm Re}\ \t\left[(C-\alpha C)^*(C-\alpha C
)\,z\right].\ee 
We need to know the linear map 
\bb   \alpha: \Omega_\dd^2\aa\longrightarrow
         \rho(\aa)+ J^2\ee  
which is determined by the two equations 
\bb
{\rm Re}\,\t\lb R^*(C-\alpha
C)\,z\rb&=&0\qquad{\rm for\ all}\ R\in\rho(\aa),
\label{a1}\\ 
{\rm Re}\, \t\lb K^*\alpha C\,z\rb
&=&0\qquad {\rm for\ all}\ K\in  J^2.\label{a2}\ee 
The solution of (\ref{a1},\,\ref{a2}) is given by 
\bb \alpha C=
\pp{\rho_L(a)&0&0&0\cr0&\rho_R(b)&0&0
\cr 0&0&\bar\rho_L^c(b,0)&0\cr 
0&0&0&\bar\rho_R^c(b,0)}+
i\pp{k\ot \Delta & 0&0&0\cr0& 0&0&0
\cr 0&0&0&0\cr 0&0&0&0}, \ee
with
\bb a=c\,\frac{\t(\Sigma'z_\ell)}{\t z_\ell - \eta\,
{\left[\t(\Delta z_\ell)\right]^2 }/{\t(\Delta^2 z_\ell)
}}, 
\qq 
    ik=-c\,\frac{\t(\Sigma'z_\ell)}{\t z_\ell- 
\eta\,
{\left[\t(\Delta z_\ell)\right]^2 }/{\t(\Delta^2 z_\ell)
}}\,
\frac{\t(\Delta z_\ell)}{\t(\Delta^2 z_\ell )}\,\eta, 
\eee
\bb
b=\frac{c_{11}\t
(M_u^*M_u)x+c_{22}\t(M_d^*M_d)x+c_{22}\t(M_e^*
M_ey)}{2Nx+\t
y+3\tilde y}, \qq c:= 
h+h^*-hh^*=1-\varphi\varphi^*.         
\eee
The Higgs potential is:
\bb V&=&\t (c^2) \left(\t\left(\Sigma'^2z_\ell\right)-
\frac{\left[\t(\Sigma'z_\ell)\right]^2}{\t z_\ell - 
\eta\,
{\left[\t(\Delta z_\ell)\right]^2 }/{\t(\Delta^2 z_\ell)
}}\right)\cr\cr  
&&+\left[|
c_{11}|^2\t(M_u^*M_u)^2+2| c_{12}|^2\t
(M_dM_d^*M_u^*M_u)+| c_{22}|^2
\t((M_d^*M_d)^2)\right]x\cr \cr 
&&+| c_{22}|^2
\t((M_e^*M_e)^2y)-\,
\frac{\left[c_{11}\t(M_u^*M_u)x+
c_{22}\t(M_d^*M_d)x+c_{22}\t(M_e^*M_ey)
\right]^2}{2Nx+\t y+3\tilde y}, \label{b1}\ee
where 
 \bb c_{11}=1-\varphi_1^*\varphi_1, \qq
c_{22}=1-\varphi_2^*\varphi_2, \qq
c_{12}=-\varphi_1^*\varphi_2, \qq
c_{21}=-\varphi_2^*\varphi_1.\ee

Let us reparameterize the scalars:
\bb h:=\pp{H+ih_Z&-h^{'*}\cr 
h&H'-ih'_Z},\qq H,H',h_Z,h'_Z\in\rr,\qq
h,h'\in\cc.         
\label{b2}\ee
 From (\ref{b1},\,\ref{b2}) we get: 
\bb V&=&4B_1H^2+4B_2H'^2+4B_3HH'
+B_4\left(| h|^2+| h'|^2-
\left(h^*h^{'}+h^{'*}h\right)\right)\cr &&
\ +\ {\rm terms \ of \  order \ 3  \ and \ 4},
\label{h}\ee where 
\bb B_1=A_1+A_2, \qq B_2=A_1+A_3, \qq
B_3=A_5, \qq B_4=2A_1+A_4; \ee
\bb A_1& =&  \t\left(\Sigma'^2z_\ell\right)-
\frac{\left[\t(\Sigma'z_\ell)\right]^2}{\t z_\ell - 
\eta\,
{\left[\t(\Delta z_\ell)\right]^2 }/{\t(\Delta^2 z_\ell)
}},\\ \cr 
 A_2&=&x\,\t(M_u^*M_u)^2-\frac{ L_1}{2Nx+\t
y+3\tilde y},\\ \cr 
 A_3&=&x\,\t(M_d^*M_d)^2+\t((M_e^*M_e)^2y)-
\frac{L_2}{2Nx+\t y+3\tilde y},\\ \cr 
  A_4&=&2x\,\t(M_u^*M_u M_dM_d^*),\\ \cr 
A_5&=&\frac{2L_3}{2Nx+\t y+3\tilde y}; \\ \cr 
 L_1&=&[x\t(M_u^*M_u)]^2, \\\cr 
 L_2&=&[x\t(M_d^*M_d)]^2+
[x\t(M_e^*M_ey)]^2+2x\t(M_d^*M_d)
\t(M_e^*M_ey),    \\ \cr 
 L_3&=&2\left(x^2\t(M_u^*M_u)\t(M_d^*M_d)+
x\t(M_u^*M_u)\t(M_e^*M_ey)\right).\ee

To get the physical variables, we  must
 diagonalize simultaneously the mass matrix
(\ref{h})~and the kinetic term in the Klein-Gordon
action. The latter  has the form:
\bb \t \left(\de\Phi^**\de\Phi\,
z\right)=
{\textstyle\frac{1}{2}}\,
 c_1\{(\partial H)^2+(\partial
h_Z)^2+
   |\partial h|^2 \}+
{\textstyle\frac{1}{2}}\,
c_2\{(\partial
H')^2+(\partial h_Z')^2+
   | \partial h'|^2 \},\ee
where 
\bb c_1=4x\,\t(M_u^*M_u), \qq 
c_2=4(x\t(M_d^*M_d)+ \t(M_e^*M_ey)).\ee
We obtain:
\bb
V=\frac{1}{2}\,m_{H_0}^2H_0^2\,+\,
\frac{1}{2}\,m_{H'_0}^2{H'}_0^2\,+\,
\frac{1}{2}\,m_{H^{\pm}}^2| H^{\pm}|^2
\,+\,{\rm terms \ of \  order \ 3  \ and \ 4}, \ee
where
\bb
H_0=\cos\theta_0\sqrt{c_1}H-
\sin\theta_0\sqrt{c_2}H',    \nonumber\\
H'_0=\sin\theta_0\sqrt{c_1}H+
\cos\theta_0\sqrt{c_2}H',    \nonumber\\
H^{\pm}=\cos\theta_1\sqrt{c_1}h-
\sin\theta_1\sqrt{c_2}h',    \nonumber\\
h_W=\sin\theta_1\sqrt{c_1}h+
\cos\theta_1\sqrt{c_2}h',    \ee
with
\bb \tan 2\theta_0=\frac{2c}{b-a}, \qq
\tan 2\theta_1=\frac{2c'}{c'_2-c'_1}, \ee 
\bb a=\frac{4B_1}{c_1}, \qq b=\frac{4B_2}{c_2}, \qq
c=-\frac{2B_3}{\sqrt {c_1c_2}}; \ee
\bb c'_1=\frac{B_4}{c_1}, \qq c'_2=\frac{B_4}{c_2},
\qq c'=\sqrt{c'_1c'_2},\ee
$\theta_0, ~\theta_1$  the Cabibbo like angles.

The masses of the Higgs  particles are given by
\bb  m_{H_0}^2=a+b+\sqrt{4c^2+(b-a)^2}, &
m_{H'_0}^2=a+b-\sqrt{4c^2+(b-a)^2},
& m_{H^{\pm}}^2=2(c'_1+c'_2), \cr 
 m_{h_{Z}}=0, & m_{h'_{Z}}=0,
& m_{h_W}=0. \ee

The masses   of  the gauge bosons 
(see Table 1 below) are found
in the covariant Klein-Gordon Lagrangian,
\bb    \t\left(\dee\Phi^**\dee\Phi\, z\right)\
 {\rm with}\ 
\dee \Phi=\de \Phi+[\rho(A)\Phi-\Phi \rho(A)]
\ee 
the covariant derivative of $\Phi$. The
normalisation of the gauge bosons is fixed by their
kinetic term in the Yang-Mills Lagrangian 
$\t \left(\rho(F)*\rho(F)\,z\right), $
with    
$ \rho(F):=\de \rho(A)+\rho(A)^2
\ \in
\Omega^2(M,\rho(\gg)).  $
$\gg$  is  the Lie algebra of the group of unitaries,
\bb\gg=u(2)\op u(1)\op u(3)
=su(3)\op su(2)\op u(1)^3.\ee
This normalisation introduces the gauge couplings
$g_i$. The Lie algebra $\gg$ being a sum of five ideals
one might expect five gauge couplings. However, the
basic object in non-commutative geometry is the
associative algebra $\aa$ which is only a sum of three
ideals,
\bb g^{-2}_3&=&\frac{4}{3}\,N\tilde x ,\\
 g_2^{-2}&=&{Nx+\t y}, \\
 g^{-2}_1&=&Nx+\frac{2}{9}N\,\tilde x+\frac{1}{2}\,
\t y+\frac{3}{2}\t \tilde y.\ee
For $g_1$ we have chosen the  gauge coupling of
the standard hypercharge.

\section{Output}

While the non-commutative version of the standard
model has one isospin doublet of scalars, the
present extension has two.
There are now four neutral scalars, two are massless,
the Goldstone bosons $h_Z$ and $h'_{Z}$ and two are
massive, the `physical' Higgs bosons $H_0$ and $H_0'$.
There are two charged scalars, the massless Goldstone
boson $h_W$ and the massive Higgs boson
$H^\pm$. The neutral and charged Higgses mix with 
Cabibbo like angles
$\theta_0$ and $\theta_1$. In the neutral sector, the
masses and the angle are
fuzzy already in the approximation $m_b\ll m_t,$ 
$m_\tau=m_c=\cdot\cdot
\cdot=0$. Note that  in this approximation $A_1=0 $.  
\bb
m_{H_0}^2=\left[2m_{t}^2+
\frac{8}{(5+z)^2}m_{b}^2\right]
\left(1-\frac{1}{6+z}\right), \ee
\bb m_{H'_0}^2=\left[2m_{b}^2-
\frac{8}{(5+z)^2}m_{b}^2\right]
\left(1-\frac{1}{6+z}\right), \ee
\bb    \tan2\theta_0=\frac{4}{5+z}
\frac{m_tm_b}{m_t^2 -m_b^2}, \ee
where we have put  $z:=\t y/x+3\tilde y/x\,>0$ and
therefore
\bb {6+z}=2(g_1^{-2}-{\textstyle\frac{1}{6}}g_3^{-2})
/x,   \label{i1}  \ee
 Let us recall the experimental values of pole masses
and gauge couplings at energies of the $Z$: 
$ m_b=4.3\pm 0.2\ {\rm GeV}, ~m_t=180\pm 12\ {\rm
GeV},~
 g_1=0.3575\pm 0.0001,\ g_2=0.6507\pm 0.0007,\ 
g_3=1.207\pm 0.026. $
Consequently $x$ ranges from 0 to $g_2^{-2}/3$ and
 $z$ ranges from 13.5 to $\infty$ and
\bb      1.38\,m_t \, < \,m_{H_0} \, <  \, 1.41\,m_t, \\  
            1.38\,m_b\,<\,m_{H_0'}\,<\,1.41\,m_b, \,\\  
            0\,<\,\sin\theta_0\,<\,0.002.    \ee   
Phenomenologically, the light Higgs $H_0'$ is a
disaster in any case. The mass of the charged Higgs
$H^\pm$ and   their Cabibbo like angle
$\theta_1$ are sharp in the above approximation,
\bb           
m_{H^\pm}&=&m_t+\frac{1}{2}\,m_b\,
\frac{m_b}{m_t},\\\cr 
   \tan2\theta_1&=&2\,\frac{m_tm_b}{m_t^2 -m_b^2},
\\ \cr 
         \sin \theta_1 &=& 0.02 . \ee
Taking into account the $\tau$ mass, however will also
render these equations fuzzy. 
\begin{center}
\begin{table} \caption{Properties of neutral gauge
bosons}
\begin{tabular}{c|cccc|c}
\renewcommand{\arraystretch}{2.5}
\\[1ex]
$\cdot$&$\gamma$&$\gamma'$&$X$&$Z'$&$Z$\\[1ex]
\hline
$g$&$g_2\sin\theta_w$&$e'$&$g_X$&$g_2$&
$g_2\cos\theta_w$\\[1ex]
$m$&0&0&$m_X$&$m_W$&$m_W/\cos\theta_w$\\[1ex]
\hline
$u_L$&2/3&$1/2-r$&1/2&1/2&
$1/2-1/6\,\tan^2\theta_w$\\[1ex]
$d_L$&-1/3&$-1/2-r$&-1/2&1/2&
$-1/2-1/6\,\tan^2\theta_w$\\[1ex]
$\nu_L$&0&0&$1/2-1/2\,\tan^2\theta_w$&1/2&
$1/2-1/2\,\tan^2\theta_w$\\[1ex]
$e_L$&-1&-1&$-1/2-1/2\,\tan^2\theta_w$&1/2&
$1/2-1/2\,\tan^2\theta_w$\\[1ex]
\hline
$u_R$&2/3&$1/2-r$&$-1/2\,\tan^2\theta_w$&0&
$-2/3\,\tan^2\theta_w$\\[1ex]
$d_R$&-1/3&$-1/2-r$&$1/2\,\tan^2\theta_w$&0&
$1/3\,\tan^2\theta_w$\\[1ex]
$e_R$&-1&-1&$\tan^2\theta_w$&0&
$\tan^2\theta_w$
\end{tabular}
\end{table}
\end{center}
Concerning the gauge bosons only the chargeless
sector is modified with respect to the standard model.
To start, we have four neutral bosons, two massless
ones, $\gamma$, the genuine photon, and $\gamma'$,
and two massive ones, $X$ and $Z'$. In the standard
model, the $Z'$ is absent and an algebraic condition
(`unimodularity') added {\it ad hoc} reduces the group
of gauge transformations $G$ by one $U(1)$ factor and
eliminates a linear combination of $\gamma'$ and $X$
leaving only the photon and the genuine $Z$. In the
standard model, the unimodularity is equivalent to 
the condition of vanishing gauge anomaly \cite{J},
\bb \t[\chi\epsilon\tilde\rho(X)^3]=0,\qq
{\rm for\ all}\ X\in\gg, \ee
where 
$ \gg:=\left\{X\in\aa\ |\ X^*+X=0\right\}$ 
is the Lie algebra of the group of unitaries $G$ and
\bb\tilde\rho(X):=\rho(X)+J\rho(X)J^{-1}\label{LA}
\ee is the Lie algebra representation that restores
invariance under charge conjugation. 
$\chi$ is the chirality operator, $\epsilon$ the
projector on the particles and $J$ the charge
conjugation. With respect to the decomposition 
$\hh=\hh_L\op\hh_R\op\hh_L^c\op\hh_R^c$ they
read:
\bb \chi=\pp{-1&0&0&0\cr 0&+1&0&0\cr
0&0&-1&0\cr  0&0&0&+1},\qq
\epsilon=\pp{1&0&0&0\cr 0&1&0&0\cr
0&0&0&0\cr  0&0&0&0},\qq 
J=\pp{0&0&1&0\cr 0&0&0&1\cr 
1&0&0&0\cr 0&1&0&0}C,\ee
where $C$ is the charge conjugation of Dirac spinors.
Table 1 recollects  the physical
properties of the neutral gauge bosons, mass, gauge
coupling and fermion charges.
We have used the following abbreviations:
\bb m_X&:=&(g_2/g_X)m_W\ \approx\  m_Z,\\\cr 
g_X^2&:=& g_2^2\,\frac{g_2^2[1-g_1^2/(6g_3^2)]}
{g_1^2+g_2^2[1-g_1^2/(6g_3^2)]}\ \approx\ 
g_2^2\cos^2\theta_w,\\\cr 
e'^2&:=& e^2\,g_1^2/({6}g_3^2) \,\cos^2\theta_w\,
\left[1-g_1^2\cos^2\theta_w/(6g_3^2)\right]^{-1}
\ \approx\ 0.011\,e^2,\\
r&:=& (g_1^{-2}+g_2^{-2})/g_3^{-2}.\ee
These approximations are good at the percent level,
$g_1^2/(6g_3^2)=0.015$.

We note that  the gauge coupling of
the $Z'$ is sharp whereas in the standard model all
 gauge couplings are fuzzy.
This sharpness comes from the fact that
$M_2(\cc)$ is simple while $U(2)$, its group of
unitaries, is not.  Phenomenologically, the $Z'$ with its
low mass {\it and} high couplings to fermions is a
disaster. On top, the $Z'$ has a gauge anomaly. We are
tempted to eliminate it with a second unimodularity
condition. Then however, its Goldstone boson remains,
another disaster.

Recent experimental evidence for deviations from the
standard model in the hadronic sector has motivated
an additional neutral gauge boson $Z'$ with a mass
around 1 TeV \cite{PC}. Clearly this $Z'$ cannot be
accommodated in the model discussed here. There
remains only one other possibility adding a $Z'$ to 
Connes' version of the standard model, $\hh\op\cc\op
M_3(\cc)$, namely to increase his algebra to 
$\hh\op\cc\op M_3(\cc)\op\cc$. Then again, it seems
impossible to have a $Z'$ mass above the top mass.

Our conclusion is that within the frame of
non-commutative geometry, it is not easy to fiddle
around the standard model.

\bigskip\noindent
It is a pleasure to acknowledge the help of Rebecca
Asquith and Lionel Carminati.

\bigskip\noindent
{\it Note added:} Alain Connes has just published a
theorem \cite{bomb} which turns his geometrical
motivation mentioned in the introduction into deep
mathematics. This theorem also unifies the standard
model with general relativity. One of the outcomes of
this unification is precisely {\it the} unimodularity,
that reduces the $X$ and $\gamma'$ to $Z$ and that
had remained unexplained so far. In this spirit, a
second unimodularity to eliminate the $Z'$ from the 
$M_3(\cc)\op M_2(\cc)\op M_1(\cc)$ model is not
available.

 \end{document}